\documentstyle[multicol,aps,prl,epsf,graphics,psfig]{revtex}

\begin{document}

\draft
\preprint{}

\title{Massive skyrmions in quantum Hall ferromagnets}

\author{M. Abolfath$^{1,2}$, Kieran Mullen$^{1}$, and H.T.C. Stoof$^{3}$} 
\address{
$^1$Department of Physics and Astronomy,
University of Oklahoma, Norman, Oklahoma 73019-0225\\
$^2$Institute for Studies in Theoretical Physics and Mathematics,
P.O. Box 19395-5531, Tehran, Iran\\
$^3$Institute for Theoretical Physics, 
University of Utrecht, Princetonplein 5, 
3584 CC Utrecht, The Netherlands}

\date{\today}

\maketitle

\begin{abstract}
We apply the theory of elasticity
to study the effects of skyrmion mass on lattice dynamics 
in quantum Hall systems. 
We find that massive Skyrme lattices behave like a Wigner crystal 
in the presence of a uniform perpendicular magnetic field. 
We make a comparison with the microscopic Hartree-Fock results
 to characterize
the mass of quantum Hall skyrmions at $\nu=1$ and  investigate how the low
temperature phase of Skyrme lattices may be affected 
by the skyrmion mass.
\leftskip 2cm
\rightskip 2cm
\end{abstract}

\def\comment#1{{$\Longrightarrow$ \sc (#1)$\Longleftarrow$ }}
\begin{multicols}{2}

\section{Introduction}
In the last few years topological spin textures
in the quantum Hall effect (QHE) have received considerable attention 
\cite{Barrett,Lee,Sondhi,Moon,Fertig,JJ,Mullen,Abolfath2,AbolfathEjtehadi,Timm,MM,Cote97}. 
The existence of skyrmions can be anticipated within the frame work of the
Chern-Simon-Landau-Ginsburg mean field theory \cite{Lee}, i.e., 
integrating out the charge fluctuations of the composite bosons yield
an effective model for the Chern-Simon's gauge field. 
The transport properties of the skyrmions can be extracted through considering
the fluctuating Chern-Simon's gauge field, which can be derived by
expanding the effective action about its minimum energy solution.
One of the leading terms of this expansion is the Maxwell action
of the Chern-Simon's fluctuating field, e.g.,
$(m^*/\bar{\rho})\sum_{k,\omega} |{\bf J}^s|^2$. $m^*$  is the electron
effective mass in a host semiconductor,
and $\bar{\rho} = 1/(2\pi \ell_0^2)$ is the average electron density
at $\nu=1$, and $\ell_0$ is the magnetic length.

Although a skyrmion mass is physically reasonable,
in the usual
minimal field theories \cite{AbolfathEjtehadi,Timm} skyrmions
are considered as massless objects.
>From the microscopic point of view, e.g., a
microscopic Hartree-Fock approximation, \cite{Fertig,Cote97}
it is not certain if the skyrmions are massive.\cite{Allan} 
The resolution of this
ambiguity between the microscopic Hartree-Fock approximation
and the Landau-Ginsburg-Chern-Simon's theory 
(considered in this paper), is an open question.

In this paper we apply the theory of elasticity
\cite{Landau} to investigate the effect of skyrmion mass on the
thermodynamic properties of a skyrmion crystal.
We derive the collective mode dispersion relations
for the Skyrme lattices at long wave lengths. 
We then make a comparison with the massless microscopic 
Hartree-Fock calculations
to reconcile the prediction of the 
Chern-Simon theory with the massless models \cite{Cote97}.
In addition, we suggest how the mass of skyrmions
may be characterized within the  microscopic picture.
We also study the stability of the Skyrme lattices at low temperature and
show that the low temperature phases of these Skyrme lattices are 
{\em not} affected 
by including a mass term in the effective action, unless if the
mass is sufficiently large.
We show the mass of skyrmions is suppressed by decreasing the Zeeman energy,
and indicating that skyrmions are massless at zero Zeeman energy.

\section{Skyrmion mass}

If skyrmions have mass, this will affect a host of properties, from their
tunneling through a constriction to the thermodynamics of a lattice of
skyrmions. 
 The rationale for their having mass is that in the starting point
for many calculations, the Chern-Simons Lagrangian of Lee and Kane,  the
electrons have mass.  After standard manipulations (introduction of a CP$^1$
field, changing variables to $\bf m$,  the local spin field) one has a
continuum theory in which gradients in the spin texture become associated
with the charge distribution.  If a skyrmion moves slowly across the
system, this corresponds to the motion of one quasiparticle.  It seems
reasonable that the motion of such a texture will involve inertial terms.
Questions similar in spirit arisen in determining the mass of
vortices \cite{thouless,ady}.

Recalling the duality relation between the topological 3-current of
skyrmions and the Chern-Simon's gauge field \cite{Mullen,Abolfath2},
$J^s_\mu = (\nu/2\pi) \epsilon_{\mu\nu\lambda} \partial_\nu A_\lambda$,
one might anticipate a second derivative term in time will appear in the
effective action.  After integrating out the statistical gauge field one
obtains in the limit of low frequency and long wavelength:\cite{Mullen}
\begin{eqnarray}
S_E[{\bf m}]& =&  {1\over 2} \sum_{{\bf k},\omega} 
\left(V(k) \left| J^s_0\right|^2 \right.  
+  {m^*\over \bar \rho} \left|{\bf J}^s\right|^2 + 
\nonumber \\ && i \alpha {\bf A}^{(0)}(-k)\cdot{\bf J}^s(k) - \nonumber \\
&&{2\pi \alpha \over k^2}
 J_0^s(-k)\, {\bf \hat z} \cdot {\bf k \times J}^s(k)
 \Bigr) \nonumber 
\\
&&+S_{NL\sigma M}+ S_z + S_{\rm fermi}
\end{eqnarray}
where the last three terms are the nonlinear sigma action, the Zeeman
contribution, and a term in the action that guarantees the skyrmions obey
fermi statistics.  The first term is electrostatic, and involves the
fourier transform of an
effective interaction potential $V(k)$ including
screening by fluctuations in the texture,
the second contains the kinetic energy and the third reflects that the 
skyrmion sees the original boson as a magnetic flux tube.  The constant
$\alpha$ is an odd integer and arises from a Chern-Simons term that makes
the system fermionic.

The skyrmion current density is related to the spin texture via:
$$
J^s_\lambda = {-\nu \over 8\pi} \epsilon_{\lambda\mu\nu}
\left(\partial_\mu {\bf m} \times \partial_\nu {\bf m} \right) \cdot {\bf
m}
$$
where the indices run over time and two spatial dimensions.  
The zeroth component is the topological charge density of the texture,
which is proportional to the quasi-particle
number density, $\rho(r) \equiv J^s_0$.
If we consider a single skyrmion
texture that moves uniformly, $J^s_\lambda=
J^s_\lambda({\bf x}- {\bf v}t)$, then it is straightforward to show 
the relation between the skyrmionic current and their charge density
satisfy the usual charge conservation law, i.e., ${\bf J}^s = \rho {\bf v}$.
Then the kinetic term simplifies to
$$
\frac{1}{2}\sum_{{\bf k}}  {m^*\over \bar\rho}  |{\bf J}^s({\bf k})|^2 
= {1\over 2} M_0 v^2
$$
which yields a transport
mass for the skyrmions $M_0 = (m^*/\bar\rho) \int d^2{\bf r} 
\rho^2({\bf r})/(2\pi)^2$. 

This mass is derived from kinetic considerations.  It is possible that the
correct ``mass''  to calculate 
will depend upon exactly what is being measured in
a given experiment. 


\section{Collective modes}

The long range order of the skyrme crystal is determined by the repulsive
Coulomb interaction, and the topological, XY interaction of the
hedge-hog fields.
An antiferromagnetic ordering between the single skyrmions, within
a square lattice minimizes the topological interaction.
This can be realized after mapping of the topological hedge-hog fields
of the charge one skyrmions onto a system of classical dipoles. 
However, a triangular lattice is favored by the Coulomb interaction,
similar to the Wigner crystals.
When the Zeeman energy is small enough, the skyrmions can pair up
into  charge
two skyrmions and lower the total energy of the lattice.
In contrast to single skyrmions where their topological 
hedge-hog fields are analog
to a system of classical dipoles, the charge two skyrmions mimic
a system of classical quadrapoles.
This favors triangular lattice ordering of charge two skyrmions, i.e.,
a bi-skyrmion lattice.

The low-lying collective modes of a Skyrme lattice consist of phonons and
spin waves.\cite{AbolfathEjtehadi,Timm,Cote97}
The dispersion relation of these collective modes can be obtained
by adding a dynamical term to the effective Hamiltonian.
This Hamiltonian has been derived in Ref.\cite{AbolfathEjtehadi}
by a first principle calculation of a non-linear $\sigma$ model,
assuming a specific skyrmion is localized and well separated from
other skyrmions, i.e., they interact weakly.
More precisely, we assume that we can divide the 2-dimensional
configuration space into $N$ regions ($N$ is the number of skyrmions)
such that a given skyrmion lives in the region that other skyrmions are close 
to their vacuum.
This condition enables us to linearize the interaction potential energy
among the isolated-skyrmions.
This assumption may break down if the size of skyrmions, $\lambda$,
becomes comparable with the distance between them.
In this case the next to the linear terms in the potential energy may be
significant, and one should take them into account.
Roughly speaking, this happens if $R \leq 2\lambda$ ($R$ is the separation
between two skyrmions).
Here the skyrmion size $\lambda$ is defined as the radius at which
the spin lies in the XY plane.
The relevant potential energy functional of the excitations is obtained by
introducing the lattice of equilibrium positions of the skyrmions $R_{i\alpha}$
is an initial lattice {\it ansatz}
($\alpha$ is the cartesian component of the position of the $i$th skyrmion), 
the displacement field $u_{\alpha}({\bf R}_i)$, and the orientation
field of the skyrmions $\theta({\bf R}_i)$. The result is then 
\begin{eqnarray}
E[{\bf u}, \theta] &=& \sum_{i \neq j}
V_0(|{\bf R}_i + {\bf u}({\bf R}_i) - {\bf R}_j - {\bf u}({\bf R}_j)|)
\nonumber\\&& + \sum_{<ij>} J(|{\bf R}_i + {\bf u}({\bf R}_i)
- {\bf R}_j - {\bf u}({\bf R}_j)|) \nonumber\\&&
\times \cos(\chi({\bf R}_i) + \theta({\bf R}_i) - \chi({\bf R}_j) 
- \theta({\bf R}_j)).
\label{Mx}
\end{eqnarray}
Here the topological hedge-hog interaction for a single-skyrmion lattice 
is given by $J(R) = c^2\tilde{g}/(4\pi^2\ell_0^2) K_0(\kappa R)$
where $\tilde{g}=ge^2/(2\epsilon\ell_0)$ is the Zeeman energy, $g$ is
the effective gyromagnetic ratio, $\kappa^2=\tilde{g}/(2\pi\ell_0^2\rho_s)$,
$\rho_s$ is the spin stiffness.
$K_0(x)$ is the modified Bessel function.
In addition, $c$ is a constant that can be obtained from the 
asymptotic form of an isolated skyrmions 
and equals $c=30.4\ell_0$. \cite{AbolfathEjtehadi}
For a bi-skyrmion lattice
$J(R) = - c^2\tilde{g}^2/(8\pi^3\ell_0^4\rho_s) K_0(\kappa R)$
and $c=79\ell_0^2$.
One should note $c$, and therefore the topological interaction
between skyrmions $J(R)$, depend on the local form of skyrmions. 

\subsection{Phonons}
We find the spectrum of the phonons,
using the standard technique  of expanding the
energy functional about its minima, i.e., ${\bf u}=0$, $\theta=0$,
assuming the orientational field of skyrmions is frozen out. 
For the single-skyrmion case $J(R)$ is positive, hence
$\chi_i - \chi_j = \pi$ is the lowest energy state of the topological
XY interaction.
For the bi-skyrmion case $J(R)$ is negative and $\chi_i - \chi_j = 0$
is the lowest energy state. 
An estimate on the total energy of skyrmions show the bi-skyrmion
configuration is likely if the Zeeman energy is very small ($\leq 10^{-5}$).
We therefore do not consider this configuration for the rest of this paper.
Expanding the potential energy in terms of the displacement fields
(up to quadratic order terms) gives
\begin{equation}
E[{\bf u}] = E_{\rm classic} + \frac{1}{2} \sum_{{\bf k}\in BZ}
\sum_{\alpha\beta} {\bf u}^*_\alpha({\bf k}) D_{\alpha\beta}({\bf k})
{\bf u}_\beta({\bf k}),
\end{equation}
with $E_{\rm classic}$ the classical groundstate energy of the Skyrme lattice
and $D_{\alpha\beta}({\bf k})$ is the dynamical matrix
\begin{eqnarray}
D_{\alpha\beta}({\bf k}) &=& \frac{2\pi}{k}\frac{e^2}{\epsilon a_c}
k_\alpha k_\beta + a_c [(\mu + \lambda) k_\alpha k_\beta 
\nonumber\\&& + \mu k^2
\delta_{\alpha\beta} + \gamma k_x k_y (1 - \delta_{\alpha\beta})
+ {\cal O}(k^4)].
\end{eqnarray}
The first term comes from the electrostatic interactions of a Wigner
crystal,\cite{Bonsall} the others arise from standard 2D elasticity theory.
The quantity $a_c$ is the area of a unit cell,
while $\mu, \lambda$, and $\gamma$ are the
conventional Lam\'e coefficients for a square lattice.\cite{Landau}

Since the effective interaction consists of two terms, i.e., the direct and 
exchange interactions, the Lam\'e coefficients can be expressed 
by means of $\mu = \mu_0 + \mu_1$,  
$\lambda = \lambda_0 + \lambda_1$ and $\gamma = \gamma_0 + \gamma_1$.
Here $0$ and $1$ are labels associated with the direct and the
exchange Coulomb energy.
Expanding the exchange term (second term) in Eq.(\ref{Mx}) up to the
quadratic order in displacement field, 
and for the single-skyrmionic square lattices with only nearest neighbor
exchange interactions we find  
\begin{mathletters}
\label{Lame}
\begin{equation}
\mu_1 = \frac{c^2 \tilde{g}}{2\pi^2 a_c} \sqrt{\frac{\pi}{2}} \;
\frac{0.5 + \kappa R}{(\kappa R)^{1/2}} e^{-\kappa R},
\end{equation}
\begin{equation}
\lambda_1 = -\frac{c^2 \tilde{g}}{2\pi^2 a_c} \sqrt{\frac{\pi}{2}} \;
\frac{7/4 + 3\kappa R + (\kappa R)^2}{(\kappa R)^{1/2}} e^{-\kappa R},
\end{equation}
\begin{equation}
\gamma_1 = \frac{c^2 \tilde{g}}{2\pi^2 a_c} \sqrt{\frac{\pi}{2}} \;
\frac{5/4 + 2\kappa R + (\kappa R)^2}{(\kappa R)^{1/2}} e^{-\kappa R}.
\end{equation}
\end{mathletters}
We also find  
the contribution of the direct interaction to the elastic constants
in units of $(e^2/\epsilon \ell_0 a_c)\sqrt{|1-\nu|/2\pi}$ are
$\mu_0=-0.146289$, $\lambda_0=-0.536199$ and $\gamma_0=-1.560224$
(we take the nearest neighbor separation between the skyrmions
$R=\ell_0\sqrt{2\pi/|\nu-1|}$ as the square lattice constant and therefore 
$a_c = R^2$). Note that $\gamma=0$ for a triangular lattices.

\begin{figure}
\center
\epsfxsize 6.0cm \rotatebox{-90}{\epsffile{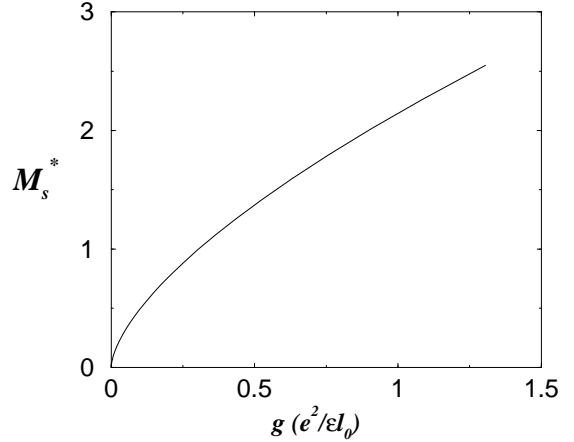}}
\vskip0.5cm
\caption{The skyrmion mass ($M_s^* \equiv 2\pi M_0/m^*
= \int d^2{\bf r} \rho^2({\bf r})$) is plotted
as a function of the $g$-factor. The Landau level filling
factor is $\nu$ and $m^*$ is the effective mass of electron in the 
semiconductor.}
\label{fig1}
\end{figure}

The time-derivative parts in the action of the displacement field
${\bf u}$, consists of the Wess-Zumino term and the kinetic energy
of the skyrmions which are first and second derivatives in
(imaginary) time, respectively. Combining these with our previous results
gives the following effective Euclidean action for the low energy
spectrum of the displacement field
\end{multicols}
\begin{equation}
S_{\rm eff}[{\bf u}] = \frac{1}{2}\sum_{\alpha\beta} \sum_{{\bf k} \in BZ}
\int_0^{\hbar\beta} d\tau \; \left[-i\varepsilon_{\alpha\beta}
m\omega_c u^*_\alpha({\bf k}, \tau) \partial_\tau
u_\beta({\bf k}, \tau)
- M_0 \delta_{\alpha\beta}
u^*_\alpha({\bf k}, \tau)\partial^2_\tau u_\beta({\bf k}, \tau)
+ u^*_\alpha({\bf k}, \tau) D_{\alpha\beta}({\bf k})
u_\alpha({\bf k}, \tau)\right],
\label{action}
\end{equation}
\begin{multicols}{2}\noindent
where 
$\omega_c$ is the cyclotron 
frequency of the electron.
The dependence of the skyrmion mass on the Zeeman energy can be obtained
by solving the non-linear differential equation for the spin texture
of a single skyrmion \cite{JJ}
and is shown in Fig. \ref{fig1}.  Note that unlike the total number of
spins participating in the skyrmion, the skyrmion mass {\it decreases} with
$g$.  This arises because the mass is proportional to the square of the
skyrmion density, and as $g$ decreases the skyrmion becomes more spread
out.
Expanding the time dependence of the complex field ${\bf u}({\bf k}, \tau)$
in terms of the bosonic Matsubara frequencies, $\omega_n = 2\pi n /\beta$,
leads to the fluctuation matrix
\begin{equation}
{\cal S}_n({\bf k}) = \left(\begin{array}{cc}
M_0\omega_n^2
+D_{xx}({\bf k}) & m\omega_c\omega_n + D_{xy}({\bf k}) \\
-m\omega_c\omega_n + D_{yx}({\bf k}) &  
M_0\omega_n^2+D_{yy}({\bf k})
\end{array}\right).
\end{equation}
The action in Eq.~(\ref{action}) is similar to the action of a Wigner
crystals in the presence of a magnetic field. Although this is as expected, it
should be noted that the magnetic field interaction is here exactly
recovered by the topological Wess-Zumino term.

\begin{figure}
\center
\epsfxsize 6.0cm \rotatebox{-90}{\epsffile{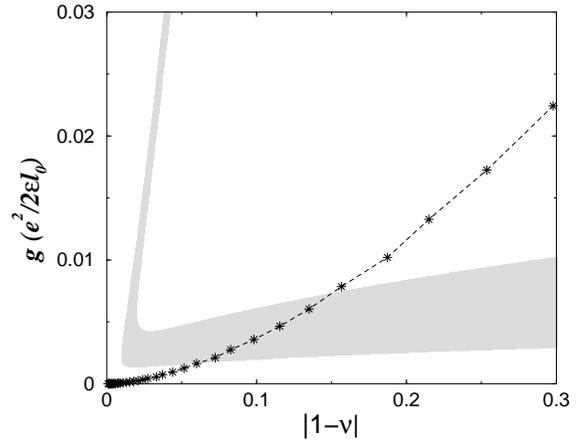}}
\vskip0.5cm
\caption{
The phase diagram for massive (and massless) skyrmions is shown.
The stable region of the square skyrmion lattices are 
represented by the points (dark region).
The triangular lattice is stable in the white region.
For $|1-\nu| \leq 0.01$, 
the square Skyrme lattices are unstable.
The curve of $R=2\lambda$ is shown. 
The overlap between skyrmions is significant below this curve 
and the topological interaction becomes strong.
}
\label{fig3}
\end{figure}

The contribution of the zero-point energy of the phonons to
the total energy of the Skyrme
lattices, may be obtained by integrating out the quadratic fluctuations in 
${\bf u}$. It turns out that $E = E_{\rm classic} + E_{\rm flu}$ where
\begin{eqnarray}
E_{\rm flu} = \lim_{\beta \rightarrow \infty} && \frac{1}{\beta} \sum_n
\sum_{{\bf k}\in BZ} \ln[\beta^2\hbar^2(\omega_-({\bf k}) + i\omega_n)
\\ \nonumber && \times
(\omega_+({\bf k}) - i\omega_n)],
\end{eqnarray}
and $\omega_\pm({\bf k})$ are the phonon frequencies
\end{multicols}
\begin{eqnarray}
\omega^2_\pm({\bf k}) = \frac{1}{2 M_0^2}
\left(m^2 \omega_c^2 + M_0(D_{xx}+D_{yy}) \pm
\sqrt{\left[m^2 \omega_c^2 + M_0(D_{xx}+D_{yy})\right]^2 
- 4 M_0^2(D_{xx}D_{yy}-D_{xy}D_{yx})}\right),
\end{eqnarray}
\begin{multicols}{2}\noindent
which have been obtained by substituting the analytical continuation
$i\omega_n \rightarrow \omega({\bf k}) + i \delta$ into the fluctuation
determinant.
The dispersion relation of the phonons consists of 
a gapped mode, as well as a gapless modes.
For the former, the gap starts at the cyclotron frequency 
$\omega({\bf k}\rightarrow 0)=(m^*/M_0)\omega_c$ 
and at long wave lengths obeys
\begin{equation}
\omega^2_+({\bf k}) = \left(\frac{m^*}{M_0}\omega_c\right)^2 + 
\frac{2\pi e^2}{\epsilon a_c M_0}k + a_c \frac{3\mu+\lambda}{M_0} k^2 
+ {\cal O}(k^3),
\end{equation}
where $\varphi$ is the azimuthal angle of the wave vector ${\bf k}$ in 
the XY plane. The gapless mode on the other hand obeys
\begin{eqnarray}
\omega^2_-({\bf k}) &=& 
\frac{2\pi e^2}{\epsilon (m^*\omega_c)^2} (\mu-\frac{\gamma}{2}\sin^22\varphi)
k^3 + {\cal O}(k^4).
\label{kkz}
\end{eqnarray}
Within a low Zeeman energy limit, where the mass of skyrmion is small, the
gapped mode goes toward the higher energies and the effect of the mass of
skyrmions becomes less significant. At this limit the prediction of 
the Chern-Simon
theory, approaches to the current prediction of the microscopic
Hartree-Fock approximation. However, for higher Zeeman energies, it is not
clear if the prediction of the microscopic Hartree-Fock approximation
leads to a single gapless mode or it can support the
gapped mode and subsequently the possibility of the existence
of a non-zero skyrmion mass too \cite{Allan}.
The square lattice can be unstable against the lattice fluctuations.
Moreover the contribution of the exchange interaction to the 
Lam\'e coefficients falls off exponentially (much faster than the 
contribution from the direct interaction) and when $\nu \rightarrow 1$
a triangular Wigner crystal configuration is more favorable than a
square lattice.
A similar situation arises when either $\tilde{g} \rightarrow 0$ or
$\tilde{g} \rightarrow \infty$. 
Our numerical studies show that this instability can be
seen for values of the filling fraction and $g$-factors that are shown in
Fig. \ref{fig3}. 
The curve of $R=2\lambda$ is also plotted.
The overlap between skyrmions is significant below this curve 
and the topological interaction becomes strong. 
The assumption of weakly interacting skyrmions may fail within
the region below this curve, and one should take the next
to the linear topological XY terms into account.
As it is shown in Fig. \ref{fig3},
for $|1-\nu| \leq 0.01$, and for certain direction of ${\bf k}$, 
the gapless mode becomes imaginary, implying an instability of the 
square 
Skyrme lattice that is again consistent with the microscopic Hartree-Fock
models \cite{Cote97}. The stable region of the square lattices is characterized
by dark area and the stable region of the triangular lattices is shown by the
white area. \cite{comment_new}
It is seen the triangular lattice reappears when the Zeeman energy 
is small enough.
Without Coulomb interactions a lattice configuration is always unstable
against the attractive interaction between skyrmions, i.e., a single
skyrmion with topological charge $N$ is the global minima of the 
skyrmionic energy functional.

As it is seen in Fig. \ref{fig1}, the quantum Hall skyrmions are massless
at $g=0$. For the massless case the gapped mode goes to infinity and
there is just only one gapless mode
\begin{eqnarray}
(m\omega_c\omega && ({\bf k}))^2 = \mu\left(
\frac{2\pi e^2}{\epsilon k} + a^2_c [\lambda + 2\mu]\right)k^4
\nonumber\\&& - 2\gamma\left(
\frac{2\pi e^2}{\epsilon k} + a^2_c [\mu + \lambda + \frac{1}{2}\gamma]\right)
k_x^2 k_y^2 + {\cal O}(k^6),
\label{kkz1}
\end{eqnarray}
which is identical to the gapless mode of the massive skyrmions.
In fact, the long wave length power-law behavior of the gapless mode is totally
unaffected by the mass of skyrmions.
For both the massless and massive theories, 
we thus find that $\omega \propto k^{3/2}$
at long wave lengths, which is
consistent with the microscopic Hartree-Fock calculations \cite{Cote97}.
Experimentally,
the mass of skyrmions may not be probed {\em directly} by phonon
excitations unless we excite the gapped modes (the cyclotron modes), e.g., by
optical measurements \cite{commm}.
Furthermore, the mass of skyrmions can not change the melting point of the
2D skyrmions, since $T_m$ is specified by the elastic constants
of Skyrme lattices (the 2D melting point just depends on
the interaction energy between skyrmions).
Returning to the evaluation of the zero-point energy of
the phonons, the dominant contribution at low temperatures is thus obtained as
$E_{\rm flu} = \sum_k \hbar\omega_-({\bf k})$. Similarly the average square
displacement of the skyrmions is given by  
\begin{eqnarray}
\langle u^2 \rangle = \ell_0^4 \sum_{{\bf k}\in BZ} &&
\frac{D_{xx}({\bf k})+D_{yy}({\bf k})}{\hbar\omega_-({\bf k})}\nonumber\\&&
\times \left[2 n_B(\beta\hbar\omega_-({\bf k})) + 1 \right],
\label{Eu}
\end{eqnarray}
where $n_B(x)$ is the Bose-Einstein distribution function.

\subsection{Magnons}
Next we consider the spin waves of the Skyrme lattices by
expanding the energy functional in Eq.~(\ref{Mx}) 
also in the orientation
field $\theta$. Up to the quadratic order, it gives the XY-energy
contribution 
\begin{eqnarray}
E[{\bf u}, \theta] = E[{\bf u}] +
\frac{1}{2} K_{XY} \int d^2{\bf r} |\nabla \theta({\bf r})|^2,
\label{Xy}
\end{eqnarray}
where $K_{XY}=J(\kappa R)$ is the effective stiffness associated with
gradients in the skyrmion orientations.
As is shown in Ref. \cite{AbolfathEjtehadi},
the effective action for the spin waves contain also a mass term
($\Lambda_0$ is the moment of inertia) and we obtain finally
\begin{eqnarray}
S_{\rm eff}[\theta] &=& \int_0^{\hbar\beta} d\tau
\int d^2{\bf r} \left(
\frac{\Delta M}{a_c}~\partial_\tau \theta 
\right. \nonumber \\ && \left.
+ \frac{\Lambda_0}{2 a_c}(\partial_\tau \theta)^2
+ \frac{K_{XY}}{2} |\nabla\theta|^2\right).
\label{Tx}
\end{eqnarray}
The first term in Eq.(\ref{Tx}) is the usual (dynamical) Berry's phase of
a quantum Hall ferromagnet, where $\Delta M$ is the average change in the total
magnetization induces by a single skyrmion texture. Note that in principle
there is also a contribution from the Hopf term in the effective action for
the quantum Hall ferromagnet, which at the quantum level ensures that the
skyrmion obeys the correct spin-statistics relation \cite{wilczek}.    
Since both these terms are a total derivative, however, the equation of motion is not
affected by these terms and 
the long wave length dispersion relation that follows from action (\ref{Tx})
turns out to be
$\omega(k) = c_s k$, where $c_s=\sqrt{a_c K_{XY}/\Lambda_0}$ is the 
velocity of the spin waves. The contribution from these fluctuations to the
total energy of the crystal is again
$E_{fl\theta} = \sum_k \hbar\omega({\bf k})$, and the mean square value of the
associated fluctuations is
\begin{equation}
\langle \theta^2 \rangle =  \hbar \sum_{{\bf k}\in BZ}
\frac{1}{\Lambda_0 \omega({\bf k})}
\left[2 n_B(\beta\hbar\omega({\bf k})) + 1\right].
\label{THT}
\end{equation}
The coupling between the displacement
and the orientational fields (${\bf u}$ and $\theta$), turn out to be 
the next to leading order terms
and are therefore negligible for our purposes.
These terms lead to interactions between
the phonons and spin waves, which we do not
consider here. As a result we find for the zero-point energy of
the phonons and spin waves simply $E_{fl} = E_{\rm flu} + E_{fl\theta}$.
>From Eq.~(\ref{Eu}), we find 
$\langle u^2 \rangle \sim |1-\nu| R^2/5$ at zero temperature.
We also find $\langle \theta^2 \rangle = \sqrt{2U/K_{XY}}$
from Eq.~(\ref{THT}) where $U = 1/(2\Lambda_0)$. 
For small values of $K_{XY}$ and/or large values of $U$
the quantum fluctuations are severe that the disordered phase can emerge.
For very small Zeeman energy, where a
phase transition from square single-skyrmion lattice
into a triangular bi-skyrmion lattice can be observed 
\cite{Cote97,AbolfathEjtehadi}, we obtain
$\langle \theta^2 \rangle \sim \tilde{g}^{0.46}$. At the limit of small
$\tilde{g}$, fluctuations are negligible and the long-range order
of the bi-skyrmion lattices is not influenced by the quantum fluctuations.

\section{Conclusion}
In this paper we have studied a system of 2-dimensional quantum Hall
skyrmions, starting from a Chern-Simon-Landau-Ginsburg mean field theory.
A Maxwell term can be obtained through the gradient expansion of the
Chern-Simon action around its minimum energy solution. This term is 
responsible for generating of the skyrmionic inertial mass. 
Away from $\nu=1$ skyrmions stay in a crystal form. The long range order
of crystal depends on the Landau level filling factor, and the Zeeman
energy. Optical phonons which are out of phase fluctuations of the skyrmion 
lattices, are gapped since skyrmions carry an inertial mass.
Therefore quantum Hall skyrmions behave like a Wigner crystal
in the presence of external magnetic field.
The inertial mass of skyrmions is vanished at zero Zeeman energy.
In this situation the optical phonons are highly gapped, and they become
inaccessible.
We finish this paper with a final comment. At $T=0$, and far from 
$\nu=1$, the zero point quantum fluctuations of the phonons destroy the
long-range order of skyrmion crystals. This is a high density limit of
skyrmions where the Coulomb interaction screens from one $1/r$ to $\ln(1/r)$
by the Chern-Simon fluctuations. In this limit, the skyrmions behave 
like a gas of classical particles, i.e., they are crystalized under
high pressures.
This yields to the possibility of observing the re-entrance of the 
solid phase, followed by the disorder (liquid) phase 
at $T=0$ when $\nu$ is far enough from $1$.

\section{acknowledgement}
We acknowledge helpful conversations with Herbert Fertig, Steve Girvin,
and Allan MacDonald.
The work at the University of Oklahoma was supported by the NSF under
grant No. EPS-9720651 and a grant from the Oklahoma State Regents for 
Higher Education.


\end{multicols}

\begin{references}
\bibitem{comment_new}
In this paper we assume the triangular lattice is the global minima
of the skyrmionic energy functional when the topological
XY interaction is negligible, in comparison with the Coulomb interaction.
However, the other possibilities of
lattice types (like the rectangular lattices) are not overruled by
symmetry, we leave them for further investigations.
\end{references}
\end{document}